\documentclass[aps,rmp,reprint,showkeys]{revtex4-1}
\usepackage{amsmath}
\usepackage{amsfonts}
\usepackage{amssymb}
\usepackage{color}
\usepackage{graphicx}
\usepackage[version=4]{mhchem}
\usepackage{natbib}
\begin{document}

\title{{Impact} of composition on the dynamics of autocatalytic sets}
\author{Alessandro Ravoni}
\affiliation{Department of Mathematics and Physics, University of Roma Tre, Via della Vasca Navale 84, 00146 Rome, Italy}

\begin{abstract}
Autocatalytic sets are sets of entities that mutually catalyse each {other's} production through chemical reactions from a basic food source. Recently, the reflexively autocatalytic and food generated theory has introduced a formal definition of autocatalytic sets which has provided {promising} results in the context of the origin of life.
However, the link between the structure of autocatalytic sets and the possibility of different long-term behaviours is still unclear.
In this work, we study how different interactions among autocatalytic sets affect the emergent dynamics.
To this aim, we develop a model in which interactions are {presented through} composition operations among networks, and the dynamics of the networks {is reproduced} via stochastic simulations. We find that
the dynamical emergence of the autocatalytic sets depends on the adopted composition operations. In particular, operations involving entities that are sources for autocatalytic sets can promote the formation of different autocatalytic subsets, opening the door to various long-term behaviours.
\end{abstract}


\maketitle

\section{Introduction}
The exact sequence of events that led to the formation of the first living organisms from non-living matter is still {a topic under debate} \cite{Luisi2016,Szostak2017,Rasmussen2004,Benner2012,Bernhardt2012}.
{On the other hand, some particular qualities of early organisms are commonly accepted and are well known.}
Among these, {we can point out the self-replication ability of the early life forms} \cite{Luisi2016,Kauffman1993,Higgs2015,Rasmussen2016,Nghe2015}.
In this scenario, the autocatalytic sets (ASs) are of great interest. {Introduced} by Kauffman \cite{Kauffman1971,Kauffman1993,Kauffman1986}, ASs are sets of entities capable of spontaneous {emergence} and {self-reproduction} through catalytic reactions, starting from a finite set of entities assumed to be available {in} the environment.
There are several definitions of ASs in the literature (see, for instance, \cite{Jain1998, Sharov1991}).
Recently, \textcite{Hordijk2011} {have introduced} the notion of reflexively autocatalytic and food generated (RAF) {sets}, a formal definition of ASs in the framework of chemical reaction systems (CRSs) 
{(see Section \ref{Background} for definitions)}.
Properties of RAF sets have been studied by various authors, and among {the most} important results (see \cite{Hordijk2017, Hordijk2018a} for {more details}) {note} the implementation of a polynomial time algorithm able to identify the presence of {a RAF set} in a general network of interacting entities \cite{Hordijk2011, Hordijk2004, Hordijk2012a} and the detection of an autocatalytic structure in the metabolic network of Escherichia Coli \cite{Sousa2015} and ancient anaerobic autotrophs \cite{Xavier2020}.

Moreover, RAF theory {has successfully proved} that RAF sets have a hierarchical structure, where the largest set of reactions within a CRS {satisfying} RAF property (the so-called maxRAF) has many subsets that are {smaller RAF sets themselves} \cite{Hordijk2018a, Hordijk2012b, Hordijk2014}.
The latter is a peculiarity that makes RAF sets potentially suitable {for experiencing} adaptive evolution (a {feature} that is generally referred to as evolvability); i.e., to collect evolutionary changes beneficial for survival and reproduction in a given environment. Indeed, it has been argued that the autocatalytic subsets present within the structure of the maxRAF could be the elementary units on which natural selection can act \cite{Hordijk2014, Vasas2012, Hordijk2018b}: the availability of spontaneous reactions would allow the occurrence of mutations and{, consequently, the} appearance of novel autocatalytic subsets able to replicate themselves with different rates and competing with each other.

However, first results show that the asymptotic dynamics of simple RAF sets eventually reaches the state in which all the reactions of the maxRAF {occur catalytically} \cite{Vasas2012, Hordijk2018b}. 
In {this} state, all the elementary autocatalytic units coexist without effectively competing {with} each other, thus leaving no room for adaptive evolution \cite{Vasas2012, Vasas2010, Hordijk2018b}.
The evolvability of RAF sets {can} be restored by embedding them into compartments and allowing the sharing of resources and the exchange of chemical molecules \cite{Vasas2012, Hordijk2018b, Kauffman2011,Serra2019}.
In fact, through numerical simulations, it has been observed that RAF sets enclosed in semipermeable protocells can reach different asymptotic states \cite{Serra2019}, {and that} spatially separated RAF sets {consuming} the same food source can give rise to different combinations of competing autocatalytic subsets \cite{Hordijk2018b}, suggesting that the evolvability of RAF sets is related to the interactions among RAF sets themselves.

In this work we investigate this latter {point}. 
In particular, we study the role of {various} interactions among RAF sets in order to understand how {these interactions affect} the emergent dynamics. 
To this aim, we use the stochastic Petri nets (SPNs) formalism \cite{Molloy1982,Haas2006} to represent and evolve RAF sets.
Furthermore, we introduce some composition operations acting on nets, which correspond to different interactions among RAF sets.
In this framework, assuming that the entire maxRAF set always emerges in an isolated RAF set, our goal is to find some composition operations under which the dynamical appearance of the maxRAF is not invariant. This means that the corresponding interaction causes only some of the maxRAF subsets to emerge, allowing the existence of multiple long-term behaviours required for the evolvability of RAF sets.

The paper is organised as follows. In Section \ref{Background} we introduce the definitions of RAF sets and SPNs. {In} Section \ref{model} {we describe} the model we use to evolve nets and {we introduce} the composition operations. In Section \ref{results} we {present and analyse} the results obtained by simulating the dynamics of {various} composed RAF sets. Finally, {in} Section \ref{conclusions} {we discuss and demonstrate the conclusions.}

\section{\label{Background}Background}
\subsection{Reflexively autocatalytic and food generated sets}
In RAF theory \cite{Hordijk2011,Hordijk2004}, a network of interacting entities is {represented} by a CRS.
{Following previous definitions \cite{Lohn1998, Hordijk2004, Hordijk2017}, we introduce a CRS as a tuple $(S, R, C)$ such that:
\renewcommand{\labelitemi}{\textendash}
\begin{itemize}
 \item $S$ is a set of entities;
 \item $R$ is a set of reactions, $\rho \rightarrow \pi$, where $\rho, \pi \in S$ are the reactants and products of a reaction, respectively;
 \item $C$ is a catalysis set, that is, a set of pairs $\{(s,r), \;s\in S\;, r\in R\}$ indicating the entity $s$ as the catalyst of reaction $r$.
\end{itemize}
We also define a food set $F \subset S$ such that entities $s \in F$ are assumed to be available from the environment.
}
{We describe a CRS as a bipartite graph such that:
\renewcommand{\labelitemi}{\textendash}
\begin{itemize}
\item nodes  are of  two kinds $V = S \cup R$;
\item edges are of two kinds $E = E_{r} \cup C$;
\item $\emptyset$ is a pseudo-entity representing the environment and $s \in S$ is a food entity if there exists a reaction $r \in R$ such that $(\emptyset,r) \in E_{r}$ and $(r,s) \in E_{r}$. 
\end{itemize}
}
{Note that} $E_{r}$ is called the spontaneous reactions set. {The edges $(s,r) \in E_{r}$ can be interpreted as source} entity $s$ is consumed by reaction $r$, while {in} edges $(r,s) \in E_{r}$ entity $s$ is produced by reaction $r$.
{$i_{s}$ denotes} a reaction such that $(\emptyset,i_{s}),(i_{s},s) \in E_{r}$, {implying} an input reaction producing a food entity. 
Moreover, we introduce outflow reactions $o_{s}$ such that $(s,o_{s}),(o_{s},\emptyset) \in E_{r}$.
Thus, the CRS is a flow reactor that allows inflow and outflow of entities from and towards the environment.

Let $R'$ {represent} a subset of $R$. The closure $cl_{R'}(F)$ is defined to be the (unique) minimal subset of $S$ that contains $F$ together with all entities that can be produced from $F$ by repeated applications of reactions in $R$. {Note that $cl_{R'}(F)$ is well defined and finite \cite{Hordijk2004}}. 
Given a CRS $(S,R,C,F)$, a RAF set is a set of reactions $R'\subseteq R$  (and associated entities) that satisfies the following properties:
\renewcommand{\labelitemi}{\textendash}
\begin{itemize}
\item Reflexively autocatalytic: for each reaction $r \in R'$ there exists at least one entity $s \in cl_{R'}(F)$ such that $(s,r) \in C$; 
\item $F$-generated: for each reaction $r \in R'$ and for each {entity} $s \in S$ such that $(s,r) \in E_{r}$, it is $s \in cl_{R'}(F)$. 
\end{itemize}
Thus, a RAF set is a set of reactions able to catalytically produce all its source entities starting from a suitable food set.
It is also possible to define the closure of a set of reactions, introducing the notion of closed RAF sets \cite{Smith2014}.
Given a CRS $(S,R,C,F)$, a subset $R' $ of $R$ is said to be a closed RAF set if: 
\renewcommand{\labelitemi}{\textendash}
\begin{itemize}
\item $R'$ is a RAF set; 
\item  $\forall \; r$ such that all its source entities and at least one catalyst are either part of the set $F$ or are produced by a reaction from $R'$, it is $r \in R'$.
\end{itemize}
{Authors of} \cite{Hordijk2018c} describe a procedure to detect closed RAF sets in a generic CRS. It has been argued that closed RAF sets are associated with the attractors of the dynamics of a CRS and therefore represent the relevant units capable of {experiencing} adaptive evolution \cite{Vasas2012, Hordijk2018b, Smith2014, Hordijk2018c}. 
\begin{figure}
\includegraphics[scale = 0.35]{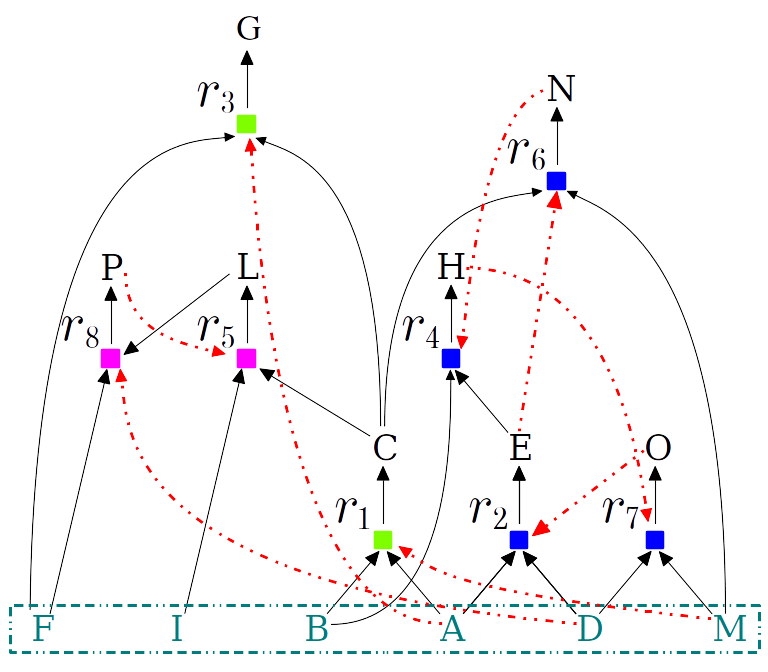}
\caption{\label{RAF_example}RAF set example. {Entities are represented by letters (dark-green for food-entities, black for non-food entities).
The food set is surrounded by a dark-green dashed rectangle.
Reactions are displayed by coloured squares.
A black arrow emerging from a letter towards a square (from a square pointing at a letter) indicates the corresponding entity is a source (a product) for that reaction. Red dashed arrows indicate catalysis.
The maxRAF set consists of three closed RAF sets: $R^{(1)}\,=\,\{r_1,r_3\}$ (green squares), $R^{(2)}\,=\,\{r_1,r_3,r_5,r_8\}$ (magenta and green squares), $R^{(2)}\,=\,\{r_1,r_3,r_2,r_4,r_6,r_7\}$ (blue and green squares).}}
\end{figure} 
FIG.~\ref{RAF_example} shows an example of a RAF set and {its constituent closed RAF sets.}
{We exemplify our point by introducing a set of entities $S = \{A,B,C,D,E,F,G,H,I,L,M,N,O,P\}$ and a food set $F = \{A,B,D,F,I,M\}$. In this example, the RAF set is composed of the following reactions (an entity above the arrow of a reaction indicates the catalyst associated with that reaction):}
\begin{align*}
 r_1&:\; \ce{A + B ->[M] C}\\
 r_2&:\; \ce{A + D ->[O] E}\\
 r_3&:\; \ce{C + F ->[A] G}\\
 r_4&:\; \ce{B + E ->[N] H}\\
 r_5&:\; \ce{C + I ->[P] L}\\
 r_6&:\; \ce{C + M ->[E] N}\\
 r_7&:\; \ce{D + M ->[H] O}\\
 r_8&:\; \ce{F + L ->[D] P}\\
\end{align*}
{The maxRAF set consists of three closed RAF sets: $R^{(1)}\,=\,\{r_1,r_3\}$, $R^{(2)}\,=\,\{r_1,r_3,r_5,r_8\}$, $R^{(2)}\,=\,\{r_1,r_3,r_2,r_4,r_6,r_7\}$.}
{Note that the closed RAF set $R^{(1)}\,=\,\{r_1,r_3\}$ is also a constructively autocatalytic and food generated (CAF) set \cite{Mossel2005}; i.e. a RAF set for which reactions can take place on the condition that their catalysts are already made available by the occurrence of catalysed reactions (starting from $F$).
More restrictive conditions of this kind result in higher rates of catalysis that form a CAF set \cite{Mossel2005}.
All the RAF sets studied in this work contain one RAF subset which in turn is a CAF set.}

\subsection{Stochastic Petri nets}
A Petri net consists of (see Petri and Reisig \cite{Petri2008} for further details):
\renewcommand{\labelitemi}{\textendash}
\begin{itemize}
\item a finite set of place $P$;
\item a finite set of transitions $W$;
\item functions $b, e : P \times W \rightarrow \mathbb{N}$.
\end{itemize}
Here $b(p,w)$ and $e(p,w)$ 
are the number of edges from place $p$ to transition $w$ and from transition $w$ to place $p$, respectively. 
The sets $b(w) \subset P$ and $e(w) \subset P$ are the sets of places connected to transition $w$ by at least one edge.
A marking $X$ of a Petri net is a map $X: P \rightarrow \mathbb{N}$ that assigns a number of tokens to each place. {In fact}, a marking $X$ identifies a state of the system in the space of possible configurations of tokens available in each place. 
{With $x_{p}$, we indicate} the number of tokens of place $p$ available in marking $X$.
Firing a transition $w$ consumes $b(p,w)$ tokens from each of its input places $p \in b(w)$, and produces $e(p',w)$ tokens in each of its output places $p' \in e(w)$.
For each marking $X$, a transition $w$ is enabled (it may fire) if there are enough tokens in its input places {making the consumption possible. This shall occur, if and only if} $X(p) \geq b(p,w)$, $\forall p \in P$. 
A stochastic Petri net \cite{Molloy1982, Haas2006} is a Petri net {for which} each transition is equipped with a (possibly marking dependent) rate for the exponentially distributed transition firing times.
$\lambda$ {denotes} the set of firing rates of a SPN.
Note that the evolution of a SPN with exponentially distributed transition rates is isomorphic to continuous-time Markov chain \cite{Molloy1982}.

{Petri net formalism provides a suitable environment for studying the composition of networks, with both a computational and theoretical approach; the latter, in particular, in the context of category theory \cite{Baez2017}. Moreover, this formalism can describe nets with different dynamics \cite{Vazquez2011}.
The authors will investigate these aspects in a forthcoming work.}

\section{\label{model}The model}
\subsection{Building the net}
Given a CRS $(S,R,C,F)$ (or a set of reactions and associated entities that is a RAF set), we build a SPN {by} adding a place $p$ for each species $s \in S$ and a transition $w$ for each reaction $r \in R$ such that $b(w) = \rho(r)$ and $e(w) = \pi(r)$, where $\rho(r)$ and $\pi(r)$ are the set of all sources and targets entities of edges $(s,r),(r,s) \in E_{r}$, respectively.
Note that with this {notation we consider} both inflowing and outflowing transitions.
Moreover, for each catalysis $(s,r) \in C$ we add a transition $w$ such that $b(w) = \rho(r) \cup s$ and $e(w) = \pi(r) \cup s$. 
Hereafter, we use $S$ and $R$ to indicate both species and reactions of a CRS and places and transitions of a SPN.
The rates $\lambda(r)$ associated {with} each transition $r$ are marking dependent rates: 
\begin{equation}
 \lambda(r) = h_{r}(X) \lambda_{r}.
\end{equation}
Here, $\lambda_{r}$ is a fixed constant depending on the type of {its corresponding} reaction in the CRS ($\lambda_{r} = \{\lambda_{s},\lambda_{c},\lambda_{i},\lambda_{o}\}$ {in which $\lambda_r$ specifies} spontaneous, catalysed, inflowing and outflowing reactions, respectively) and $h_{r}(M)$ is a value proportional to the number of combinations of tokens available in the input places of transition $r$ in the state $X$.
{Thus, explicitly, we shall have:}
\begin{equation}
 h_{r}(X) = \frac{\prod_{j}b(j,r)}{V^{|b(r)|-1}} \prod_{j}\binom{x_{j}}{b(j,r)},
\end{equation}
where the product is among all the input places of transition $r$.
We set functions $b$ and $e$ such that the inflowing transitions do not consume tokens of the pseudo-entity $\emptyset$ and produce a fixed value of tokens of the food entities, while the outflowing transitions consume a token of the outflowing entities and do not produce tokens of the pseudo-entity $\emptyset$.
Thus, the rate of outflowing transitions is proportional to the amount of tokens of the outflowing entity, while the rate of inflowing transitions is independent of the state of the system.
It is {noteworthy} that the inflowing of food elements still remains a stochastic event.
{We add inflowing transitions for entities not belonging to the originary food set $F$ (setting the rate of such transition equal to zero), if required for the purpose of composition between nets (see Section \ref{sec_comp}).}
The dynamics of the obtained SPN is described by the stochastic mass action kinetics, that is the {classical} dynamics used to represent chemical reactions, assuming a well-stirred system \cite{Anderson2011}.

The CRSs used in this work are generated according to the binary polymer model (BPM) \cite{Kauffman1986}.
The BPM produces a CRS where the entities set $S$ consists of all bit strings up to (and including) a maximum length $N$, and the reaction set $R$ consists of condensation and cleavage reactions. {Condensation reaction is a concatenation of two bit strings resulting in a longer string, and cleavage reaction cats a bit string into two smaller ones.}
The food set is represented by all entities with {a} length less than or equal to a fixed length $l_{f}$ (we set $l_{f} = 2$), and each entity can be a catalyst of each reaction with a certain probability fixed a priori.

We chose to {only allow the condensation reactions to occur in the system.}
{Note that the technique we use is also applicable in the case where cleavage reactions are {allowed}. In fact, none of the operations we introduce in Section \ref{sec_comp} is affected by the reversibility of the network reactions.
Including cleavage reactions would produce networks, in principle, with different dynamics, since a network with irreversible reactions can have different topology with respect the equivalent reversible network \cite{Feinberg1995}.
The authors are currently investigating this aspect in an upcoming work.}

{With} this limitation, all the spontaneous reactions of our model are binary reactions (i.e., reactions with two reactants, possibly of the same entity type), while all the catalysed reactions are ternary reactions.
Even if ternary reactions {are rare}, they can represent a first approximation of two or more elemental reactions, such as the sequence of reactions of an enzyme catalysis \cite{Gillespie2007}.

\subsection{\label{sec_comp}Composition}
We model interactions between CRSs as composition operations between SPNs.
First, note that RAF sets satisfy the following conditions (hereafter, we refer to {these} as the inclusion conditions) \cite{Hordijk2004}:
\renewcommand{\labelitemi}{\textbullet}
\begin{itemize}
\item if $R_1'$ is RAF in $(S_1,R_1,C_1,F_1)$, it is RAF also in $(S_2,R_2,C_2,F_2)$, if conditions $S_1 \subseteq S_2, \; R_1 \subseteq R_2, \; C_1 \subseteq C_2, \; F_1 \subseteq F_2$ are satisfied;
\item if $R_1'$ is RAF $(S_1,R_1,C_1,F_1)$ and $R_2'$ is RAF in $(S_2,R_2,C_2,F_2)$, then $R_1' \cup R_2'$ is RAF in $(S_1 \cup S_2,R_1 \cup R_2,C_1 \cup C_2,F_1 \cup F_2)$.
\end{itemize}
Thus, if composition does not remove entities from the food set or reactions belonging to a RAF set, it {will not have any impact} on its RAF property. However, the dynamical behaviour of the composed system can be, generally, different from that of the starting one.

Let $(S_{1},R_{1},b_{1},e_{1},\lambda_{1})$ and $(S_{2},R_{2},b_{2},e_{2},\lambda_{2})$ be two SPNs and let $I,O$ be subsets of their inflowing and outflowing transitions sets.
Let $\sim$ be the equivalence relation such that $s \sim s'$ if $i_{s} \in I$ and $o_{s'} \in O$ for some choice of $I,O$.
$S_{\sim}$ {denotes} the set of places identified by relation $\sim$.
We define the following composition operations:
\renewcommand{\labelenumi}{$\mathrm{CO_{\Roman{enumi}}}$:}
\renewcommand{\labelitemi}{}
\begin{enumerate}
\item \label{COI}\begin{itemize}
        \item $(S_{1},S_{2}) \rightarrow S_{*} = S_{1} \cup S_{2}$;
        \item $(R_{1},R_{2}) \rightarrow R_{*} = (R_{1} \cup R_{2} \cup R_{\Roman{enumi}}) \setminus (O_{1} \cup I_{2})$;
        \item $(b_{1},b_{2}) \rightarrow b_{*} = b_{1} \cup b_{2} \cup b_{\Roman{enumi}}$;
        \item $(e_{1},e_{2}) \rightarrow e_{*} = e_{1} \cup e_{2} \cup e_{\Roman{enumi}}$;
        \item $(\lambda_{1},\lambda_{2}) \rightarrow \lambda_{*} = \lambda_{1} \cup \lambda_{2} \cup \lambda_{\Roman{enumi}}$;
        \item $R_{\Roman{enumi}} := \{r\, |\, b(r) = s,\,e(r) = s', \; \forall \;  s \in S_{1},s' \in S_{2} \text{ such that } s \sim s'\}$;
        \item $\lambda_{\Roman{enumi}} := \{\lambda(r)\, |\, \lambda(r) = h_{r}(X) \lambda_{f} , \; \forall \; r \in R_{\Roman{enumi}}\}$.
       \end{itemize} 
 \item  \label{COII}\begin{itemize}
        \item $(S_{1},S_{2}) \rightarrow S_{*} = S_{1} \cup S_{2}$;
        \item $(R_{1},R_{2}) \rightarrow R_{*} = (R_{1} \cup R_{2} \cup R_{\Roman{enumi}}) \setminus (O_{1} \cup I_{2})$;
         \item $(b_{1},b_{2}) \rightarrow b_{*} = b_{1} \cup b_{2} \cup b_{\Roman{enumi}}$;
        \item $(e_{1},e_{2}) \rightarrow e_{*} = e_{1} \cup e_{2} \cup e_{\Roman{enumi}}$;
        \item $(\lambda_{1},\lambda_{2}) \rightarrow \lambda_{*} = \lambda_{1} \cup \lambda_{2} \cup \lambda_{\Roman{enumi}}$;
        \item $R_{\Roman{enumi}} := \{r\, |\, b(r) = b(r'),\,e(r) = b(r''), \; \forall \;  r'' \in R_{2} \text{ such that } b(r'') \subset S_{\sim}\text{ and } b(r')\sim b(r'')\}$;
        \item $\lambda_{\Roman{enumi}} := \{\lambda(r)\, |\, \lambda(r) = h_{r}(X) \lambda_{f} , \; \forall \; r \in R_{\Roman{enumi}}\}$.
       \end{itemize}
 \item  \label{COIII}\begin{itemize}
        \item $(S_{1},S_{2}) \rightarrow S_{*} = \{S_{1} \sqcup S_{2}\}/\sim$;
        \item $(R_{1},R_{2}) \rightarrow R_{*} = R_{1} \cup R_{2}$;
        \item $(b_{1},b_{2}) \rightarrow b_{*} = b_{1} \cup b_{2}$;
        \item $(e_{1},e_{2}) \rightarrow e_{*} = e_{1} \cup e_{2}$;
        \item $(\lambda_{1},\lambda_{2}) \rightarrow \lambda_{*} = \lambda_{1} \cup \lambda_{2}$.
       \end{itemize} 
\end{enumerate}
Here $\lambda_{f}$ is a constant value and $(S_{*},R_{*},b_{*},e_{*},\lambda_{*})$ is the composed SPN.

To summarise, all the composition operations we define relate a set of places $S$ that are input for outflowing transitions of a SPN, together with a set of places $S'$ that are input for inflowing transitions of another SPN{. The} formal addition of inflowing transitions for places not belonging to the food set {enlarges} the possible composition operations between SPNs.

{Given a set $S_{\sim}$, operation $\mathrm{CO_{I}}$ adds a transition from $S$ to $S'$ for each pair of places in $S_{\sim}$, while operation $\mathrm{CO_{II}}$ adds a transition from $S$ to $S'$ for each combination of places that appears as input of a transition in the inflowing net.
Each combination corresponds to the definition of complex \footnote{Note that complexes play a major role in the framework of chemical reaction networks theory. For instance, the deficiency theorems
\cite{Feinberg1995} are able to predict whether the dynamics of a large class of networks will have a stationary distribution, starting from the topology of the reaction graph having complexes as nodes.} in the framework of chemical reaction networks \cite{Feinberg1995}. In fact, given a set of chemical species, a complex is defined as a member of the vector space generated by the species that provides the inputs (or the outputs) of a reaction \cite{Feinberg1995, Anderson2011}.}

{Both operation $\mathrm{CO_{I}}$ and $\mathrm{CO_{II}}$ introduce a flux of entities from one net to another. The composite network can therefore be seen as the union of two separate networks that evolve in parallel, communicating only with (asymmetrical) exchange of chemical species. This could be, for instance, the case of two spatially separated protocells, one of which can release molecules towards the other. According to this interpretation, the flowing rate $\lambda_f$ is a parameter that encompasses the characteristics of the flow process (for example, cell permeability). Operation $\mathrm{CO_{III}}$, instead, merges each pair of places in $S_{\sim}$, allowing transitions of the two original SPNs to operate on the glued set of places. Composing nets via operation $\mathrm{CO_{III}}$ actually produces a new single network. In this case, one can think of composite net as the result of mutations that enlarge a network (for instance, net $1$), introducing new possible reactions and, consequently, new chemical species (corresponding to net $2$).}

It is worth to underline that, if {there exists a} $s \in S_{\sim}$ such that $s \in F$, operations $\mathrm{CO_{I}}$ and $\mathrm{CO_{II}}$ can actually modify the RAF property of net $2$.
In particular, if transitions $r \in R_{(I,II)}$ are assumed to be spontaneous transitions ($\lambda_{f} < \lambda_{c}$), net $2$ could {not be} catalytically produced starting from the food set $F$. { In this case,} the composed net contains a RAF set $R'$ such that $R_{1} \subseteq R' \subset (R_{1} \cup R_{2})$, with $R_{1} = R'$ if $F \subseteq S_{\sim}$. 
Instead, if transitions $r \in R_{(I,II)}$ are assumed to be (auto) catalysed transitions ($\lambda_{f} \geq \lambda_{c}$), the whole composed net {shall be} a RAF set.

\section{\label{results}Results and Discussion}
In this Section, we present the results regarding the impact of composition on the dynamics of RAF sets {as  {follows}: we first introduce the characteristics of simulations and the quantities taken into account.
In Section \ref{results_iso} we present results for the non interacting nets, in order to have a reference model for the interacting cases, presented in Section \ref{results_comp}.}

Starting from different instances of the BPM with $N < 8$, we use the RAF algorithm introduced in \cite{Hordijk2011} in order to detect and select three different RAF sets, each of which contains more than one closed RAF set.
Note that, even if different RAF sets have the same species set $S$, the set of reactions $E$ {will} be different; i.e., {various} RAF sets have different chemistry.
The RAF sets identified through this procedure constitute the collection on which we will carry out the study.
Even if such a small collection cannot be taken as a solid statistical basis, it {is still sufficient for providing us with} interesting information.
We duplicate each RAF set and let it interacts with its copy by switching to representation of RAF set as an SPN and using one of the composition operations introduced above.
We simulated the dynamics of the system using the standard Gillespie algorithm \cite{Gillespie1976, Gillespie1977}, setting the volume of the system at $V = 1$ (arbitrary units).
For each simulation, we perform $100$ independent runs of $10^6$ time steps.
One of the {necessary conditions for} a RAF set is the ability to produce itself starting from the elements of the food set.
{Indeed}, the initial state of the SPN is set such that:
\begin{equation}\label{initial_condition}
\begin{cases}    
x_{s}(t = 0) = x_{0} \text{ if $s \in F$};\\
x_{s}(t = 0) = 0 \text{ otherwise}.
\end{cases}
\end{equation}
Here $x_{0}$ is an arbitrary constant.
The {values} of $x_{0}$, $\lambda_{c}$, $\lambda_{i}$ and $\lambda_{o}$ are set such that the number of tokens of food places at $t \rightarrow \infty$ {is equal to $10^5$, for an SPN with inflow, outflow and all (and only) binary transitions having, as input, food places only (and firing rate $\lambda_{c}$).}
The value of $\lambda_{s}$ is fixed at $\lambda_{s} = \lambda_{c}/10$, while $\lambda_{f}$ varies such that $\lambda_{f} \in [\lambda_{c} 10^{-1}, \lambda_{c} 10^6]$.
{It is noteworthy that the values of these parameters are not taken from ``in vivo'' data,
but {they have phenomenological motivations.} Thus, although we can reasonably generalise the characteristics of the dynamics, quantities such as the species' production rate or the time evolution of the concentrations may differ from those in other similar stochastic simulations \cite{Hordijk2012a, Hordijk2018b}.}

We focus our attention on the effective appearance of a maxRAF set during the evolution of the system. 
In particular, we introduce the following quantities:
\renewcommand{\labelenumi}{\arabic{enumi})}
\begin{enumerate}
 \item $M_{s}(t) = \sum_{s} \frac{1}{1+x_{s}(t)} $, $\forall s \notin F$\label{prova};
 \item $\tau_{i} = \text{min}\{t\,|\,n(r) \geq i$, $\forall r \in R\}$.
\end{enumerate}
Here $n(r)$ is the number of executions of transition $r$, and $R$ is the maxRAF set.
{Note that the natural condition $x_s(t) \geq 0$ {implies} that each term contributing to the computation of $M_{s}(t)$ can possess, at most, the value one.}
Both $M_{s}(t)$ and $\tau_{i}$ are calculated for each single net that forms the composed net.
Let $m_M = \overline{M_{s}(t)}$ be the mean value of $M_{s}(t)$ for $t \rightarrow \infty$. 
If all the non-food entities of a maxRAF set are efficiently produced, then $M_s(t) \rightarrow 0$ for $t \rightarrow \infty$ and $m_M = 0$.
However, the definition of a RAF set does not ensure that all the entities associated with such a set are present in large amount during the evolution of the system.
For instance, a non-food entity that is a source for a transition of a RAF set could be continuously consumed by that transition as soon as it is produced, resulting in a fluctuating evolution of its number of tokens.
{For this reason, we consider a less restrictive condition than $m_M = 0$ for the emergence of a maxRAF set.
In particular, {for large $t$, we require a strictly positive concentration for all non-food entities and a high concentration for most of them.}
This implies that the relation {$\frac{1}{1+x_s}<1$ holds for all the terms contributing to the calculation of $M_{s}$, and that the relation $\frac{1}{1+x_s} \approx 0$ holds for most of them.}
Thus, we introduce the following condition:
\begin{equation}
m_{M} < 1. \label{RAF_condition}
\end{equation}
We assume that, if condition (\ref{RAF_condition}) holds, the entire maxRAF set emerges.

Note, moreover, that different growing rates among entities of a RAF set correspond to different effective firing rates of the transitions.
Therefore, even if a maxRAF $R$ set appears, the time $\tau$ necessary to perform all the transition of $R$ can exhibit different {slopes} during the evolution of the net, {based on the various} effective rates of subsets of $R$.
We use the slope $m_{\tau}$ of the straight line $y_{\tau}(i) = m_{\tau} i + q_{\tau}$ that approximates $\tau_{i}$ for $i \rightarrow \infty$ in order to compare the efficiency of the (total) self-production of different RAF sets.
We summarise the results obtained from the simulated composition operations and the different RAF sets of our collection through a scatter plot showing the values $(m_{M},m_{\tau})$ averaged over independent runs (FIG.~\ref{Scatter}).

\subsection{\label{results_iso}Non interacting nets}
\begin{figure}
\includegraphics[scale = 0.38]{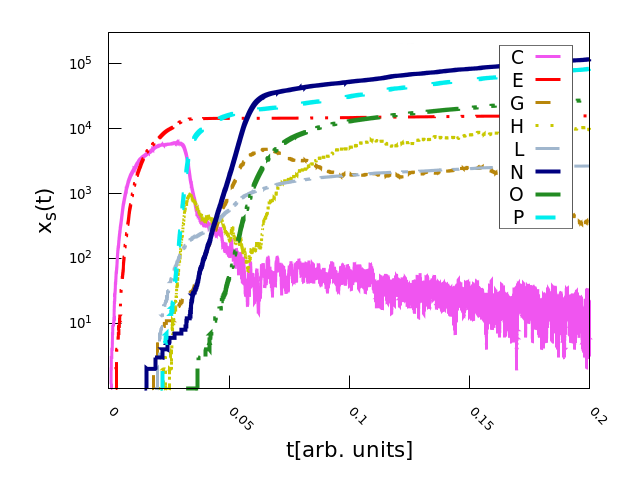}
\caption{\label{run}Isolated RAF set. Number of tokens over time of non-food entities obtained in a simulation run of the isolated RAF set shown in FIG.~\ref{RAF_example}.
After a transient time of $t \approx 0.006$ (arbitrary units) all the entities are present in a large amount, with the exception of the entity $C$ which is repeatedly consumed by more than one reaction of the RAF set.}
\end{figure}
\begin{figure}
\includegraphics[scale = 0.38]{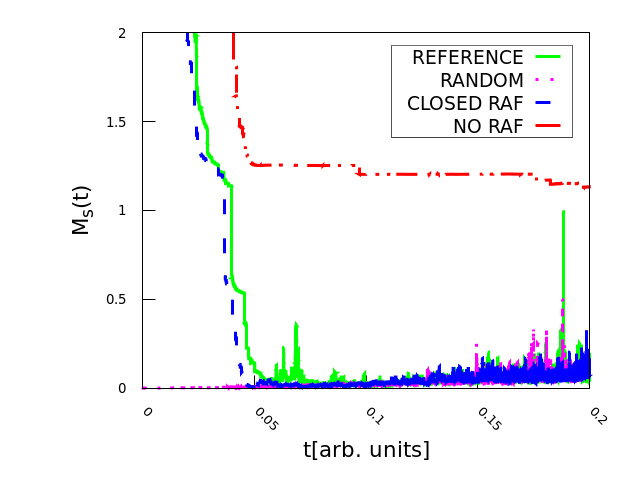}
\caption{\label{iso_m}Total production of non-food entities by isolated nets. $M_{s}(t)$ obtained in a simulation run of the isolated RAF set shown in FIG.~\ref{RAF_example} for different initial conditions: solid green line {(reference simulation)}: $x_s(t = 0) = 100$ if $s \in F$, $x_s(t = 0) = 0$ otherwise;
dotted magenta line: $x_s(t = 0) = 100 $ if $s \in F$, $x_s(t = 0) = rand(0,10^4)$ otherwise;
dashed blue line:
$x_s(t = 0) = 100 $ if $s \in F$, $x_s(t = 0) = 10^4$ if $s \in \{E,H,N,O\}$, $x_s(t=0) = 0$ otherwise.
After a transient time of $t \approx 0.006$ (arbitrary units) $M_s(t)$ always {takes} values close to zero, indicating that the maxRAF set has emerged for all the different tested configurations. 
Dash-dotted red line: net that failed to be a RAF set, obtained by {switching} off the catalysis $(D,8)$ in the RAF set shown in FIG.~\ref{RAF_example}.
In this case, $M_s(t)$ is always greater than $1$, indicating that at least one non-food {entity} is not produced by transitions of the system.}
\end{figure}
\begin{figure}
\includegraphics[scale = 0.38]{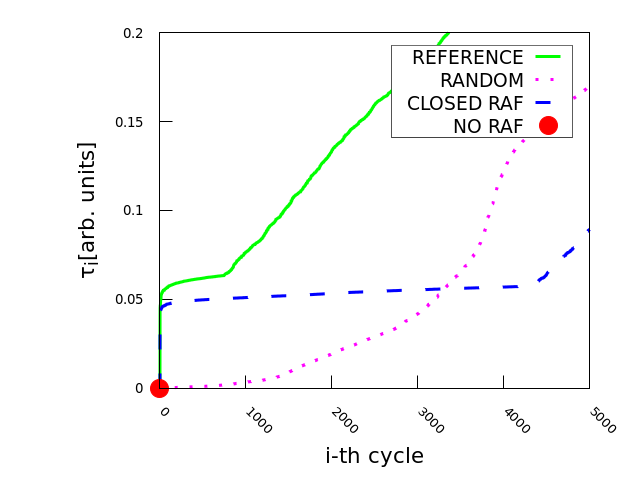}
\caption{\label{iso_t}Cycles of transitions completed by isolated nets. $\tau_{i}$ obtained {for} the isolated RAF set shown in FIG.~\ref{RAF_example} {with} different initial conditions: solid green line {(reference simulation)}: $x_s(t = 0) = 100$ if $s \in F$, $x_s(t = 0) = 0$ otherwise;
dotted magenta line: $x_s(t = 0) = 100 $ if $s \in F$, $x_s(t = 0) = rand(0,10^4)$ otherwise;
dashed blue line:
$x_s(t = 0) = 100 $ if $s \in F$, $x_s(t = 0) = 10^4$ if $s \in \{E,H,N,O\}$, $x_s(t=0) = 0$ otherwise.
The similar slope of $\tau_{i}$ for $t \rightarrow \infty$ {associated with} different tested configurations indicates that, after a transient time, the efficiency {of performing} all the transitions of the RAF set does not depend on the initial conditions.
Red circle: {the} net that failed to be a RAF set, obtained by {switching} off the catalysis $(D,8)$ in the RAF set shown in FIG.~\ref{RAF_example}.
In this case, the net is not able to perform all its transitions.}
\end{figure}
We start {investigating the non interacting nets} by simulating the evolution of three isolated RAF sets that constitute the collection, in order to obtain the dynamics that will be the benchmark for the evolution of the composite nets. 
FIG.~\ref{run} shows the time evolution of the number of tokens of non-food entities for the isolated RAF set represented in FIG.~\ref{RAF_example}.
{Hereafter, we refer to this simulation as ``reference'', since it will be used as a benchmark for the other simulations.}

It is evident that, after a transient time of approximately $0.06$ time units, all the entities associated with the RAF set grow in number, with the exception of entity $C$: $C$ is the only entity in the set to be a source for more than one transition (FIG.~\ref{RAF_example}).
{The} green line in FIG.~\ref{iso_m} shows the corresponding trend of $M_{s}(t)$.
As expected, after the same transient time of $\approx 0.06$ time units, $M_{s}(t)$ decreases down to values close to zero, the entire maxRAF set appears and condition (\ref{RAF_condition}) is satisfied (FIG.~\ref{Scatter}). 

We find that the asymptotic dynamics of simple isolated RAF sets is not affected by changing the initial state. {This is obvious from FIG.~\ref{iso_m} and FIG.~\ref{iso_t} by comparing the trends of $M_{s}(t)$ and $\tau_{i}$ obtained for} the same RAF set for {various} initial conditions.
In particular, we perform different simulations by setting $x_s(t = 0)$ equal to a random number less than $10^4$ for all non-food entities of the maxRAF set, and by setting $x_s(t=0) = 10^4$ for only {those} entities associated with a particular closed RAF set.
For all the RAF sets in the collection, the resulting values of $m_M$ and $m_{\tau}$ are in agreement with those corresponding to the initial conditions described by Eq. (\ref{initial_condition}) (FIG.~\ref{Scatter}). 
Conversely, the dynamics emerging in a net that failed to be a RAF set is significantly different (FIG.~\ref{iso_m} and FIG.~\ref{iso_t}, red lines).

These results suggest once again that simple RAF sets have an effective advantage in self-reproduction over non-RAF set{. Also,} the structure of RAF sets alone is not sufficient to guarantee the presence and the dynamical selectability of different long-term behaviours.
Using the dynamics of isolated RAF sets as reference, we can now move on to the dynamics of composed RAF sets.

\subsection{\label{results_comp}Composite nets}
\begin{figure}
\includegraphics[scale = 0.38]{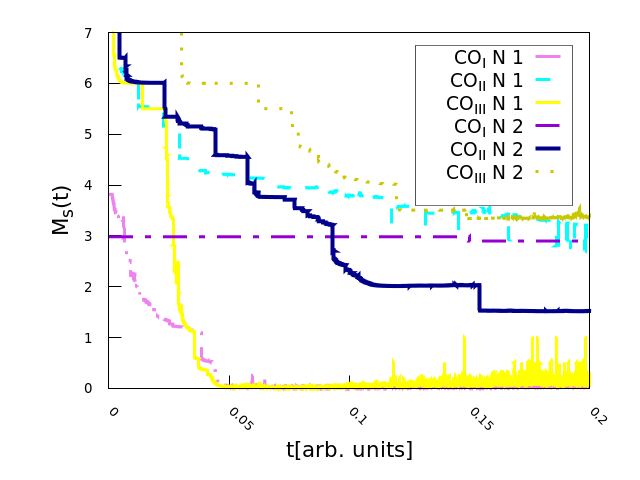}
\caption{\label{comp_m}Total production of non-food entities by composite nets. $M_{s}(t)$ {is} obtained {for} two copies of the RAF set shown in FIG.~\ref{RAF_example} {with} different composition operations.
For each composite net, two $M_s(t)$ are calculated, each from the entities associated with the {original} copies.
Dash-dotted magenta line: operation $CO_I$, net $1$, $\lambda_f = 10^5 \lambda_c$.
Dashed cyan line: operation $CO_{II}$, net $1$, $\lambda_f = 10^5 \lambda_c$.
Solid yellow line: operation $CO_{III}$, net $1$.
Dash-dotted violet line: operation $CO_I$, net $2$, $\lambda_f = 10^5 \lambda_c$.
Solid blue line: operation $CO_{II}$, net $2$, $\lambda_f = 10^5 \lambda_c$.
Dotted dark-yellow line: operation $CO_{II}$, net $2$, delayed.
For all composite operations, $M_s(t)$ of net $2$ does not satisfy {the condition stated by} Eq. (\ref{RAF_condition}) {that results in the emergence} of the maxRAF set.
For operation $CO_{II}$, {both the maxRAF sets of net $1$ and net $2$ do not emerge.}}
\end{figure}
\begin{figure}
\includegraphics[scale = 0.38]{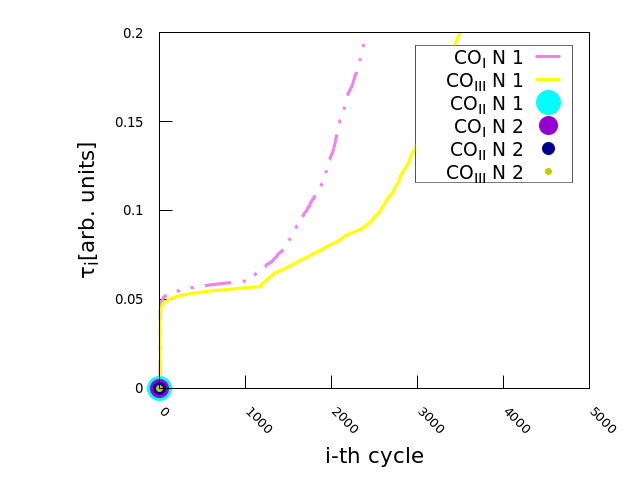}
\caption{\label{comp_t}Cycles of transitions completed by composite nets. $\tau_{i}$ {is obtained for} two copies of the RAF set shown in FIG.~\ref{RAF_example} {with} different composition operations.
For each composite net, two $\tau_{i}$ are calculated, each from the transitions associated with the textcolor{red}{original} copies.
Dash-dotted magenta line: operation $CO_I$, net $1$, $\lambda_f = 10^5 \lambda_c$.
Solid yellow line: operation $CO_{III}$, net $1$.
Large cyan circle: operation $CO_{II}$, net $1$, $\lambda_f = 10^5 \lambda_c$.
Intermediate violet circle: operation $CO_I$, net $2$, $\lambda_f = 10^5 \lambda_c$.
Intermediate blue circle: operation $CO_{II}$, net $2$, $\lambda_f = 10^5 \lambda_c$.
Small dark-yellow circle: operation $CO_{II}$, net $2$, delayed.
For all composite operations, net $2$ is not able to {execute} all its transitions.
For operation $CO_{II}$, {both net $1$ and net $2$ do not perform a complete cycle of transition of the RAF set.}}
\end{figure}
In order to {compose nets}, we choose five different sets $S_{\sim}$: the set of places belonging to the food set, the set of places non belonging to the food set, the set of places corresponding to the molecules of length $l = l_{f} +1 $ {and $l \leq l_{f} +1$} of the BPM and the set of places that are not input places for spontaneous transitions (not selected for operation $\mathrm{CO_{II}}$). 
We compose copies of the RAF sets according to the composition operations $CO_I$, $CO_{II}$ and $CO_{III}$.
The initial states of the nets are set according to Eq. (\ref{initial_condition}).
Moreover, for operation $CO_{III}$, {simulations are performed in which the transitions belonging to net $2$ cannot proceed for a certain time interval of $10^4$ time steps.}
Hereafter, we refer to this particular configuration as the ``delayed configuration''.

We find that composition operations do not have any impact on the emergence of the maxRAF sets for any choice of $S_{\sim}$ that does not include food entities (FIG.~\ref{Scatter}). However, a primordial form of biological interactions can be established{; namely}, facilitation and cheating.
In particular, in nets composed through operations $CO_I$ and $CO_{II}$, inflowing of external entities can facilitate the appearance and the sustenance of a RAF set, improving its production efficiency. At the same time, {withdrawing} entities produced by a RAF set can counteract its production.
These aspects are well highlighted by {various trends of the generated $m_{\tau}$ due to different conditions.}

On the other hand, as expected, composition operations involving the food set have a major role on the emergence of the maxRAF set.
FIG.~\ref{comp_m} and FIG.~\ref{comp_t} show the {behaviour} of $M_{s}(t)$ and $\tau_{i}$ {associated with} two RAF sets that are composed by operations $CO_I$, $CO_{II}$ and $CO_{III}$ for $S_{\sim} = F$. Similar trends are obtained from the composition of the other RAF sets in our collection.
Results show that operations $CO_I$ and $CO_{II}$ prevent the entire maxRAF set {in} at least one of the two copies from emerging.
In particular, we find that, if the rate $\lambda_{f}$ of transitions allowing the flux of food elements from net $1$ to net $2$ is low enough ($\lambda_f < 10^6 \lambda_c$), operation $CO_I$ prevents the emergence of the maxRAF set {in} net $2$, while net $1$ exhibits the same dynamics of the isolated net, as can be {concluded} from FIG.~\ref{Scatter}. 
{On the other hand}, for the same composition operation, if the rate $\lambda_f$ is sufficiently high ($\lambda_f \geq 10^6 \lambda_c$), the flow of food elements is such that the maxRAF set of net $1$ does not contain available resources to perform all its transitions, while net $2$ evolves as if it {is} isolated and able to drawn food directly from the environment (FIG.~\ref{Scatter}).
{For operation $CO_{II}$, we have achieved significantly different results.} In this case, we find different threshold values $\lambda_f = (\lambda_{f1},\lambda_{f2})$ (depending on the specific RAF set of the collection) such that the {emergence} of the maxRAF set {in} net $2$ is prevent for $\lambda_f \leq \lambda_{f1}$, while the opposite situation is obtained for $\lambda_f \geq \lambda_{f2}$.
Moreover, for intermediate values $ \lambda_{f1} < \lambda_f < \lambda_{f2}$, {no maxRAF set emerges but only some of the closed RAF sets} (FIG.~\ref{comp_m} and FIG.~\ref{comp_t}).
Therefore, for this range of values, the flux of  entities is such that both nets $1$ and $2$ have enough food elements to fire transitions and perform (complementary) subsets of the maxRAF set, {namely}, the closed RAF sets.
Furthermore, the stochastic nature of the flow process allows the emergence of different closed RAF sets in each run, hence showing the possible selectability of asymptotic dynamics for composite nets.

The observed dynamics lead to some important considerations: first, it is clear that the actual availability of resources is a crucial element in the dynamical realisation of a RAF set, and the rate at which food elements enter the net is as relevant as the definition of the food set itself{, as expected.}
Moreover, the {differences emerging due to} the impact of the $CO_I $ and $CO_{II}$ operations suggest that the complexes play an important role in the dynamics of the RAF sets, even if they are defined starting from the single entities.
We will {investigate these points} in a following work.
Finally, as previously observed in \cite{Hordijk2018b}, biological interactions different from competition among RAF sets are {plausible}.

{It is also intriguing} that an effective competition can emerge if two {nets} share the same food, as in case of composition operation $CO_{III}$ and $S_{\sim}$.
In particular, we observed that in the delayed configuration, the presence of the maxRAF of net $1$ prevents the emergence of the maxRAF of net $2$.
{In fact}, once the maxRAF of net $1$ has had enough time to appear, the number of tokens of its associated entities increases.
Since the effective rate of a transition is proportional to the number of tokens of its sources, entities of net $1$ have {an higher chance} of reacting with respect to {their counterparts} belonging to the delayed net $2$.
Most of the food elements are therefore consumed by transitions of net $1$. Once activated, only some of the transitions of net $2$ are able to be performed efficiently, leading to the emergence of only some of the closed RAF sets which constitute the maxRAF set of net $2$.
Different runs show that the emerging closed RAF sets can {vary} due to the stochastic nature of the system evolution, thus guaranteeing the selectability of the different long-term behaviours.

This result is in contrast with the previous ones where it has been observed that isolated RAF sets are not able to experiment different asymptotic dynamics.
{In fact}, composing two RAF sets using operation $CO_{III}$ {produces} a composite net in which all transitions form a (larger) RAF set ({see the} inclusion conditions, Section \ref{sec_comp}), and the effect of the delay can be seen as a selection of a particular initial state.
However, the same composite net is not able to {produce} competition if the delay is not introduced{. Also,} an effective competition between closed RAF sets {has not been} observed in an isolated net {with} initial conditions {containing RAF sets already emerged} at time $t = 0$ (FIG.~\ref{Scatter}, FIG.~\ref{iso_m} and FIG.~\ref{iso_t}).

We suggest that {key elements} for this form of competition are both the structure of the composite RAF set and the particular choice of initial conditions.
{In fact, in nets} composed by operation $CO_{III}$ and $S_{\sim} = F$, each transition that consumes at least one food entity as a source or a catalyst, always has at least one competitor {represented by} its copy.
{By} contrast, the hierarchical structure of RAF sets does not guarantee such level of competition between different close RAF sets.
Moreover, the delay brings the system into a state that {promotes} the formation of some subsets of the RAF set{. The system can hardly reach this state} only through random fluctuations.
The results presented in this paper show that these two conditions allow RAF sets to have different accessible asymptotic dynamics.
\begin{figure}
\includegraphics[scale = 0.4]{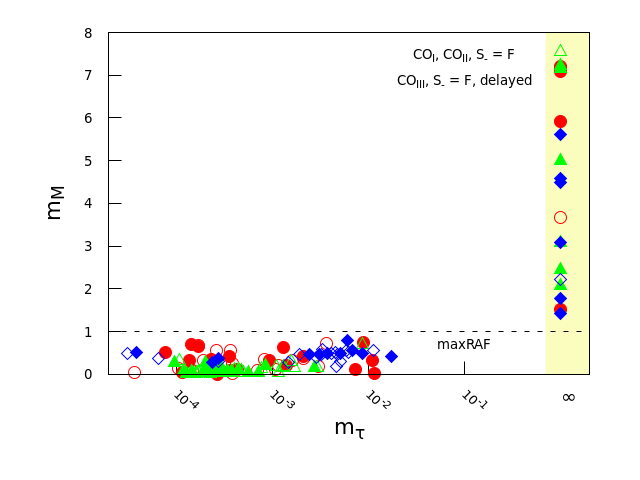}
\caption{\label{Scatter}Impact of composition operations on the emergence of the maxRAF set. On the $x$-axis: slope $m_\tau$ of the straight line {fitting} $\tau_i$ for $t \rightarrow \infty$, averaged over $100$ independent runs.
On the $y$-axis: $m_M = \overline{M_s(t)}$, averaged over $100$ independent runs. Here $\overline{\cdots}$ is the average over time for $t \rightarrow \infty$.
For each run, both $m_\tau$ and $m_M$ are calculated for $t \geq \frac{5}{6} t_f$, where $t_f$ {signifies the end of the simulation run.}
Green triangles: RAF set $1$ (Fig.~\ref{RAF_example}).
Red circles: RAF set $2$.
Blue squares: RAF set $3$.
Empty points: net $1$.
Filled points: net $2$.
Yellow shaded region indicates values corresponding to $m_{\tau} \rightarrow \infty$, obtained for composition operations $CO_{I}, CO_{II}, CO_{III}$ (delayed) and $S_{\sim} = F$.}
\end{figure}

\section{Conclusions}\label{conclusions}
In this work we study the impact of composition operations on the dynamics of simple RAF sets{. This allows us} to test whether the interactions among RAF sets {permit} different long-term dynamical behaviour of the RAF sets themselves, that is a necessary condition for evolvability. 
To this aim, we {generate} {various} RAF sets starting from different instances of the BPM and represent RAF sets as SPNs.
Moreover, we introduce composition operations that, acting on SPNs, correspond to interactions among RAF sets.
We find that, if the composition operations do not involve the food entities of a RAF set, the dynamics of the system always reaches {a} state in which all the composed RAF sets appear.
However, {how fast} the RAF sets emerge and their efficiency in self-reproduction depends on the exchange of entities, showing that the composition operations can give rise to interactions with ecological features.

On the other hand, composition operations involving food entities {hinders} the appearance of, at least, one of the two original RAF sets, giving rise to possible different long-term behaviours.
In particular, if the food entities can be exchanged between one RAF set and another, as the case of composition operations $CO_I$ and $CO_{II}$, the emergence of the maxRAF set of the starting nets depends on the rate {of} the exchange transitions (the flow).
For sufficiently low rates, only the RAF set capable of acquiring food directly from the environment is able to form. {Conversely}, for sufficiently high rates the elements of the food are exchanged in {such portions to} allow only the receiving RAF set to appear.
Furthermore, if complexes of a reaction are involved in the exchange and not individual entities (as in the case of the composition operation $CO_{II}$), we find an intermediate interval of flow rate {values within which} the exchange of food elements between the nets allows the emergence of just some of the closed RAF sets that constitute the starting maxRAF sets. For these {intermediate flow rates}, therefore, {the necessary evolvability conditions are met.}
Finally, we find that only some closed RAF sets within a maxRAF set can appear if the maxRAF set shares the food entities with its copy that has already fully emerged.
The latter is a relevant result, since sharing the same food set by two RAF sets produces a maxRAF set that is the union of the two starting RAF sets. The dynamics observed in this case {shows} that different long-term behaviours are possible for a single RAF set, at least as long as the system is in a particular initial state and the subsets of the RAF set compete with each other for each reaction that needs food elements.

In previous works it was theorised that separate RAF networks could compete for shared resources \cite{Hordijk2014, Vasas2012, Kauffman2011}, and a competitive dynamics was observed in spatially separate RAF sets \cite{Hordijk2018b}. 
Results presented in this paper confirm the possible evolvability of a system of RAF sets separated into compartments. {We also noticed that isolated (composite) RAF sets can experience different asymptotic dynamics.}
Furthermore, recent results show the presence of RAF sets in real biological systems \cite{Sousa2015, Xavier2020}, confirming their biological interest.
However, since all the simple {(tested)} isolated RAF sets experience the emergence of the entire maxRAF set, the definition of the RAF sets does not seem to be sufficient {for implying} dynamics with multiple selectable long-term behaviours. 

In order to further clarify this last point, we will explore a larger ensemble of RAF sets in a forthcoming paper.
Moreover, in the future it might be interesting to study the composition of RAF sets enclosed in protocells, investigating how the coupling between internal networks and boundaries affects the global dynamics of the system.

\begin{acknowledgments}
We {would like} to thank Marco Pedicini for his ideas, insightful help and valuable discussions. 
We want to acknowledge helpful suggestions from Marco Villani
and Fatemeh Zahara Majidi, the latter also for improving the manuscript with a precise and accurate proof-reading.
\end{acknowledgments}

\end{document}